\documentclass[twocolumn,pra,aps,amssymb]{revtex4}

\usepackage{epsfig}

\begin{document}

\newcommand{\wobie}{\leftrightharpoons}
\newcommand{\tr}{{\rm tr}}

\def\com#1{{\tt [\hskip.5cm #1 \hskip.5cm ]}}
\def\be{\begin{equation}}
\def\ee{\end{equation}}
\def\ben{\begin{eqnarray}}
\def\een{\end{eqnarray}}
\def\>{\rangle}
\def\<{\langle}
\def\hcal{{\cal H}}
\def\bcal{{\cal B}}

\title{Rates of asymptotic  
entanglement transformations for bipartite mixed states: Maximally entangled states are not special}

\author{Micha{\l} Horodecki,
Aditi Sen(De), and Ujjwal Sen }
\affiliation{Institute of
Theoretical Physics and Astrophysics, University of Gda\'{n}sk, 80-952
Gda\'{n}sk, Poland}

\begin {abstract}
We investigate the asymptotic rates of entanglement transformations
 for bipartite mixed states by local operations and classical communication (LOCC). We 
analyse the relations between the rates for different transitions and obtain simple
lower and upper bound for these transitions. 
In a transition from one mixed state to another and back, the amount of 
irreversibility can be different for different target states. Thus in a natural 
way, we get the concept of ``amount" of irreversibility in asymptotic 
manipulations of entanglement. We investigate the behaviour of these 
transformation rates for different target states.
We show that with respect to asymptotic transition rates under LOCC, the maximally entangled states do not have a special status. 
In the process, we obtain that the 
entanglement of formation is additive for all maximally correlated states. 
This allows us to show irreversibility in asymptotic entanglement manipulations 
for maximally correlated states
in \(2\otimes 2\).
We show that the possible 
nonequality of distillable entanglement under LOCC and that under operations
preserving the positivity of partial transposition, is related to the behaviour of 
the transitions (under LOCC)
to  separable target states.
\end{abstract}


\maketitle

\section{Introduction}

\label{intro}

Investigations into the emerging science of quantum information
has led to the widespread belief that entanglement in states shared 
between two systems can be used as a resource in nonclassical applications \cite{book1, book2}.
It is important to stress that such applications are independent of what 
interpretation one chooses of the Hilbert space formalism of quantum mechanics
and it keeps itself clear of the paradoxes that entanglement has been a storehouse of. 
Given a state shared between two partners, traditionally called Alice
and Bob, we will therefore like to know whether it is 
possible to use it in some communication task, for example in quantum 
teleportation \cite{tele}. 
However a given state may not immediately lend itself for use
in the envisioned communication task. One 
may have to transform it to another state, suitable for the particular communication task.
And since the state is shared between two partners, there will be natural restrictions on 
the allowed operations on the state, in the sense that Alice and Bob will be able to act 
on the state only locally. It turns out that it is useful to allow them to share
information over a classical channel also. Entanglement being a resource, they will like 
to do such transformations optimally.

Suppose therefore that Alice and Bob share the state \(\rho\), while for 
their communication task, they require the state \(\sigma\). Let 
\(R(\rho \rightarrow \sigma)\) be the optimal asymptotic 
rate at which this transformation occurs faithfully, under
local operations and classical communication (LOCC) between the sharing parties. Throughout this 
paper, \(\rho \rightarrow \sigma\) will imply a transition of a bipartite state \(\rho\) into a bipartite state 
\(\sigma\) under LOCC.

A fundamental question is whether Alice and Bob lose anything irreversibly 
during this transformation.
That is, if Alice and Bob now tries to retrieve the state \(\rho\), do they lose anything
during the cycle from \(\rho\) to \(\rho\) via \(\sigma\). More precisely, do we have 
\begin{equation}
\label{definition}
\rho \wobie \sigma \equiv R(\rho \rightarrow \sigma)R(\sigma \rightarrow \rho)  
\end{equation}
equal to unity? We will call the  quantity \(\rho \wobie \sigma\) as the \emph{amount} of 
the irreversibility in the transition
\(\rho \to \sigma \to \rho\).

In quantum information, the maximally entangled states have a special significance. For
example, they are 
the only states which one can use in faithful teleportation. It was convenient therefore 
to have special names for the rate \(R(\rho \rightarrow \sigma)\) and 
the inverse of the rate 
\(R(\sigma \rightarrow \rho)\), when \(\sigma\) is a maximally entangled state 
in \(2 \otimes 2\) \cite{tensorname, 2x2, concpure}. They are respectively called the distillable entanglement
\(D(\rho)\) and entanglement cost \(F(\rho)\) of \(\rho\) \cite{concpure, huge, Rains, HHT}.

The above question has been answered in the case when \(\sigma\) is a maximally 
entangled state in \(2 \otimes 2\) \cite{irrev, DVC}. In these 
references, there are states exhibited for which distillable entanglement is strictly less 
than its entanglement cost \cite{bound, PRLPLA}. That is, examples of \(\rho\) 
were given for which 
\(\rho \wobie \sigma < 1\), with  \(\sigma\) being a  maximally entangled state in \(2 \otimes 2\).

There is therefore a possible irreversible loss of entanglement when one transforms
\(\rho\) into a  maximally entangled state. Returning back to \(\rho\),
we may not be able to get back the entanglement with which we had started with. 
A question related to the above question can now be asked. How will the \emph{amount} of 
the irreversibilities, for such return trips of \(\rho\), change for different target
states \(\sigma\)? 
Stated in terms of the rates defined above, for given \(\rho\), how does 
\(\rho \wobie \sigma\) behave for  
different \(\sigma\)?  
This question is the main theme of this paper.

%There are many other interesting questions 
%that may arise in the context of mixed states conversions.

In this paper we analyse the conversion rates $R(\rho\to \sigma)$ 
themselves, as well as the relations between rates for different 
transitions.
In Section \ref{def}, we give simple upper and lower bounds for the optimal rate 
in a cycle of \(\rho\) to \(\sigma\) and back in terms of 
asymptotic entanglement measures. 
In Section \ref{cont}, we show that \(\rho \wobie \sigma\) is 
continuous when \(\rho\) and \(\sigma\) remains in an open set of distillable states \cite{Vidalcont}.
 In section \ref{Bellmixtures}, we will consider irreversibility
in transformations of \(\rho\) to \(\sigma\) and back in the case when \(\rho\) and \(\sigma\) are 
mixtures of two  Bell states.
In section \ref{maxnotspecial}, we show that the cycle \(\rho \rightarrow
\sigma \rightarrow \rho\) may have varying degrees of irreversibility depending on 
the chosen \(\sigma\). And there is nothing special for a cycle of \(\rho\) via 
a  maximally entangled state.
The irreversibility 
\(\rho \wobie \sigma\)
of \(\rho\) in a cycle via \(\sigma\) can be 
\emph{strictly} greater or less than its irreversibility in a 
cycle via a  maximally entangled state. %(in \(2 \otimes 2\)).
In Section \ref{maxcorr}, we show along the lines of Ref. \cite{DVC} (see also 
\cite{Shor}) that the entanglement of formation 
is additive for the maximally correlated 
states \(\sum a_{ij}\left|ii\right\rangle\left\langle jj\right|\) in arbitrary dimensions and 
this leads to the computation of the entanglement cost of such states in \(2 \otimes 2\)
 \cite{Wootters}.
For maximally correlated states in arbitrary dimensions, we 
express their entanglement cost as a simple optimization procedure.
As a by-product, we obtain irreversibility 
(with respect to a maximally entangled state) 
in asymptotic manipulations for these states in 
\(2 \otimes 2\). Using the  value of  entanglement cost of maximally correlated states (in \(2 \otimes 2\)), we will 
discuss in Section \ref{samemaxcorr},
 that the feature of  
nonextremal nature of maximally entangled states (as  studied in %\ref{Bellmixtures} and 
Section \ref{maxnotspecial}) can also be obtained by considering the class of 
%if the initial state is a 
maximally correlated states in \(2 \otimes 2\).
 We subsequently show in Section \ref{ratio} that these considerations 
can also be seen from the perspective of the ratio problem of entanglement measures 
\cite{Nielsen}. And the ratio problem discussed in this paper, is in a sense complementary 
to the one considered in
Ref. \cite{Nielsen}. In the last section (Section \ref{conclu}), 
we present some discussions.
Distillable entanglement of a bipartite state under LOCC is no greater 
than that under operations preserving the positivity of partial transposition 
\cite{PeresHorodecki, Rainssemi}. Whether a strict inequality holds is unknown.
In the concluding section, we show that such nonequality is related to the behaviour of 
the transformation rates (under LOCC) to  separable target states.

\section{Definitions and some bounds}
\label{def}

%Before giving the definition of \(\rho\) to \(\sigma\) and back and its upper bounds, l
Let
us first fix our notations. We use both the notations 
\(R(\rho \rightarrow \sigma)\) and \(D_{\sigma}(\rho)\) for the 
\emph{optimal} rate of the transformation \(\rho \rightarrow \sigma\)
for bipartite states \(\rho\) and \(\sigma\) under LOCC. 
Similarly we use 
\(1/R(\sigma \rightarrow \rho)\) and \(F_{\sigma}(\rho)\)
interchangeably.  
It will be also useful to introduce a notation for rates of transitions 
of the general form
\[
\rho_1 \to \rho_2 \to\ldots \to \rho_n \to \ldots \to \rho_2\to \rho_1.
\]
The rate of such transitions is the {\it optimal} ratio of final number copies of the state 
$\rho_1$ to the initial number of copies of $\rho_1$, \emph{in the route specified above}, and it will 
be denoted by 
\[
\rho_1 \wobie \rho_2 \wobie \ldots \wobie \rho_n.
\]
We have 
\be
(\rho\wobie\sigma)= R(\rho\to\sigma) R(\sigma \to \rho) = 
{D_\sigma(\rho) \over  F_\sigma(\rho)}
\ee
(we will sometimes put brackets, in order not to confuse between ``$=$'' and  
``$\wobie$'').
By definition we obtain
\be
(\rho\wobie\sigma)=(\sigma\wobie\rho)
\ee
and 
\be
(\rho \wobie \sigma \wobie \omega) = (\rho \wobie \sigma) 
(\sigma \wobie\omega).
\label{eq-id}
\ee
Since there are more possibilities for going from $\rho$ to $\omega$
directly, rather than via the intermediate state $\sigma$ we obtain 
the inequality 
\be
\rho \wobie \sigma \wobie \omega \leq \rho\wobie \omega.
\label{eq-ineq}
\ee
The above properties will help us to establish some bounds for
$\rho\wobie\sigma$ in terms of  the quantities  $\rho \wobie \psi^{-}$ 
and $\sigma \wobie \psi^{-}$, where we take the singlet
\[\left|\psi^{-}\right\rangle = \frac{1}{\sqrt{2}}(\left|01\right\rangle - \left|10\right\rangle)\]
 as our  canonical maximally entangled state in \(2 \otimes 2\).
(We will use \(\psi^{-}\) to denote  
\(\left|\psi^{-}\right\rangle \left\langle \psi^{-}\right|\).)
First, due to  (\ref{eq-ineq}), we have 
\[
\psi^{-} \wobie \rho\wobie \sigma \leq  \psi^{-} \wobie \sigma,
\]
which, in view of (\ref{eq-id}), gives 
\be
\rho\wobie\sigma \leq {\sigma \wobie \psi^{-} \over \rho \wobie \psi^{-}}. 
\label{eq-ineq2}
\ee
On the other hand, from  (\ref{eq-ineq}) we can also get
\be
\rho \wobie \psi^{-}\wobie \sigma \leq \rho\wobie \sigma.
\label{eq-ineq3}
\ee
Joining relations (\ref{eq-ineq2}) and (\ref{eq-ineq3}) 
and exchanging the roles of $\rho$ and $\sigma$  in (\ref{eq-ineq})
we finally obtain 
\begin{equation}
\begin{array}{rcl}
&(\rho\wobie \psi^{-})(\sigma\wobie \psi^{-}) \leq & \\
&(\rho  \wobie  \sigma) &\\ 
&\leq \min \left\{ {\sigma \wobie \psi^{-} \over \rho \wobie \psi^{-}}, 
{\rho \wobie \psi^{-} \over \sigma \wobie \psi^{-}}   \right\}&.
\end{array}
\label{bounds}
\end{equation}

So far we have related the operational quantity $\rho\wobie\sigma$
to other operational quantities (rates involving singlets).
There arises therefore the question whether one can improve the inequalities by 
use of the results on asymptotic entanglement monotones 
\cite{miary,Donald}.
For example, it is known that entanglement measures
satisfying some assumptions give upper bounds for $D$ (distillable entanglement) and 
lower bounds  for $F$ (entanglement cost).  
Thus to obtain suitable bounds, one does not have to go into the very difficult issue 
of optimizing distillation or formation task, but rather one can
choose a function with the needed properties.
On the other hand, to have for example a \emph{lower} bound 
for $D$, an  operational approach usually cannot be overcome:
one needs to point out a specific protocol of distillation.
And similarly for obtaining an \emph{upper} bound of entanglement cost. 
It turns out that in our case, there is a similar issue. We will be 
able to prove an upper bound that refers to entanglement  monotones,
rather than to conversion rates.  To this end let us recall \cite{review} 
that  if a function $E$ is (i) nonincreasing under LOCC, 
and (ii) asymptotically continuous, then we have 
\be
\label{khub}
R(\rho\to\sigma) \leq {E^\infty(\rho)\over E^\infty (\sigma)}
\ee 
where  $E^\infty(\rho)=\lim_{n\to\infty} {1\over n} E(\rho^{\otimes n})$
is the {\it regularization} of $E$. Note that the conditions are on the function \(E\), while the bound 
is with its regularization. 
Examples of such measures are entanglement of formation  and relative entropy distance from 
separable states \cite{Plenio} or the so called PPT states   
\cite{Rainssemi}. Now, recalling that 
$\rho\wobie\sigma= R(\rho\to\sigma) R(\sigma\to\rho)$ and 
applying the inequality (\ref{khub}) with different entanglement 
measures to each factor we obtain
\begin{equation}\label{fromentmeas}
\rho\wobie\sigma \leq {E_1(\rho) E_2(\sigma) \over E_2(\rho) 
E_1(\sigma)},
\end{equation}
where $E_i$ are regularizations of any two chosen  measures
satisfying (i) and (ii). Putting $E_1=F$, $E_2=D$ we 
can recover the right-hand-side bound of formula  (\ref{eq-ineq2}) (even though 
we do not know  if $D$ satisfies (ii) or if it is a regularization of some measure
which satisfies (i) and (ii)).
It can be written by means of $F$ and $D$ as 
\begin{equation}
\label{important}
\rho\wobie \sigma \leq \frac {F(\rho)D(\sigma)}{F(\sigma)D(\rho)}.
\end{equation}
Note that eq. (\ref{important}) (which is the same as eq. (\ref{eq-ineq2})) is \emph{not} obtained 
from eq. (\ref{fromentmeas}). It is obtained (as shown just before eq. (\ref{eq-ineq2}))
from general considerations on the rates of transformations.

To make the results transparent, we introduce the following quantity:
\begin{equation}
\label{diff}
R_{Diff}= R_{Diff}(\rho, \sigma) = (\rho \wobie \psi^-) - (\rho \wobie \sigma),
\end{equation}
whose continuity properties we will consider next.

\section{Continuity}
\label{cont}

In this section we present some continuity arguments for \(R_{Diff}\) that we will use later on. 
This follows from the results in 
 Ref. \cite{Vidalcont}. 
The only requirement that is needed to be imposed on an asymptotic 
measure of entanglement \(E\), for it to be continuous \emph{in any open set of distillable states},
is that 
\[E(\eta_1) \geq \frac{y}{x}E(\eta_2)\] 
whenever the transformation 
\[x \times \eta_{1} \rightarrow y \times \eta_{2} \quad (x, y \geq 0)\]  
is achievable in the asymptotic limit
by LOCC for two bipartite states \(\eta_1\) and \(\eta_2\). 
Here by \(x \times \eta\), we mean \(x\) copies of \(\eta\), with suitable 
changes when \(x\) is not a positive integer.  
We now show that \(F_{\sigma}(.)\)
satisfies this condition.

By definition of \(F_{\sigma}(.)\), \(x \times \eta_{1} \rightarrow 
y \times \eta_{2}\) implies 
\[
\begin{array}{rcl}
\displaystyle
& \quad & F_{\sigma}(\eta_1) \times \sigma 
\rightarrow \eta_1 \rightarrow \frac{y}{x}\eta_2 \\ 
& \Leftrightarrow & \frac{x}{y}F_{\sigma}(\eta_1) \times \sigma 
\rightarrow \eta_2 \\
& \Leftrightarrow & F_{\sigma}(\eta_2) \leq \frac{x}{y}F_{\sigma}(\eta_1) \\ 
& \mbox{i.e.} & F_{\sigma}(\eta_1) \geq \frac{y}{x}F_{\sigma}(\eta_2).
\end{array}
\]
The proof that \(D_{\sigma}(.)\) also satisfies the condition required for continuity, is 
similar. And therefore we have the continuity of \(R_{Diff}\) for \emph{arbitrary} 
\(\rho\) and \(\sigma\) in any
open set of distillable states. We stress that this proof is 
%not only for two Bell mixtures, but 
for arbitrary \(\rho\) and \(\sigma\) in arbitrary dimensions. We will 
 use this continuity later on, to
understand the nature of \(R_{Diff}\) in general.

\section{Irreversibility of the cycle \(\rho \to \sigma \to \rho\) for different \(\sigma\)}
\label{Bellmixtures}

We will now use the bounds obtained in Section \ref{def}, to tackle the problem 
of irreversibility of the cycle \(\rho \to \sigma \to \rho\) for 
different \(\sigma\). 
We will like to ask as to when the strict inequality of the following form 
is possible:
\begin{equation}
\label{main}
\rho \wobie \sigma < 
\rho \wobie \psi^{-}.
\end{equation}
 Definitely it is the case for states for which we can show 
\begin{equation}
\label{gerakol}
\frac {D(\sigma)}{F(\sigma)} < \left(\frac {D(\rho)}{F(\rho)}\right)^2. 
\end{equation}
This follows from inequality (\ref{important}).
%for two states \(\rho\) and \(\sigma\). 
Below we show that the inequality (\ref{gerakol}) is indeed satisfied for 
some choices of  \(\rho\) and \(\sigma\) as mixtures of two Bell states. 
More examples will be reported in Section \ref{samemaxcorr}.

%In this section, we will give an (implicit) examples 
%for which the inequality (\ref{main}) is satisfied for \emph{unequal} 
%\(\rho\) and \(\sigma\). 
We take \(\rho\) and \(\sigma\) as  mixtures of two Bell states with different mixing parameters.
Let 
\begin{equation}\label{rhotwoBell}
\rho = (1-p)\left|\phi^{+}\right\rangle\left\langle \phi^{+} \right|
+ p\left|\phi^{-}\right\rangle\left\langle \phi^{-} \right|, \quad p\in[\frac{1}{2},1]
\end{equation}
where
\[
\begin{array}{rcl}
\displaystyle \left| \phi^{+}\right\rangle  & = & {\frac{1}{\sqrt{2}}\left( \left| 00\right\rangle +\left| 11\right\rangle \right) },\\
\left| \phi^{-}\right\rangle  & = & {\frac{1}{\sqrt{2}}\left( \left| 00\right\rangle -\left| 11\right\rangle \right) }
\end{array}
\]
And let \(\sigma\) be another mixture of the same Bell states:
\begin{equation}\label{sigmatwoBell}
\sigma = (1-q)\left|\phi^{+}\right\rangle\left\langle \phi^{+} \right|
+ q\left|\phi^{-}\right\rangle\left\langle \phi^{-} \right|, \quad q\in[\frac{1}{2},1].
\end{equation}

\begin{figure}[tbp]
  \epsfig{figure=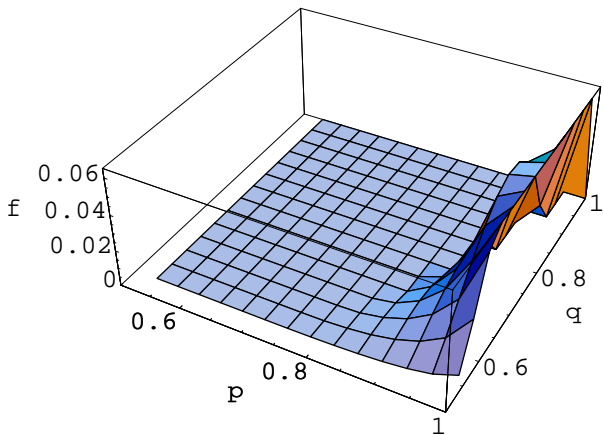,width=0.45\textwidth} 
\caption{Plot 
indicating nonzero value of \(R_{Diff}\). The function 
\(f = D^2(\rho)F(\sigma) - F^2(\rho)D(\sigma)\), with \(\rho\) and 
\(\sigma\) given by  eqs. (\ref{rhotwoBell}) and (\ref{sigmatwoBell}), is plotted
on 
the \((p, q)\)-plane. A positive value of \(f\) indicates
a positive value of \(R_{Diff}\).} \label{nonzerof}
\end{figure}

In this case, we know the values of \(D\) and \(F\) exactly \cite{DVC}, and in
certain regions on the \((p, q)\)-plane, the inequality (\ref{gerakol})
is satisfied. In Fig. \ref{nonzerof}, we plot the function 
\[f = D^2(\rho)F(\sigma) - F^2(\rho)D(\sigma)\]
over the \((p, q)\)-plane. For any \(q\), there exists a nonzero range of \(p\) near \(p = 1\),
for which \(f\) is positive. The inequality (\ref{gerakol}) is therefore satisfied in those regions of the 
\((p, q)\)-plane.
 Consequently, in those regions of the \((p, q)\)-plane, 
we have 
\((\rho \wobie \sigma) < 
(\rho \wobie \psi^-) 
\). 
Note that this region cannot have a nonempty intersection with the line
\(p=q\). We will discuss in the next section (Section \ref{maxnotspecial}), 
the inequality (\ref{main}) cannot be satisfied in this case.

\section{There's nothing special about a singlet}
\label{maxnotspecial}

As singlets are useful in many quantum information processes, one may tend to believe 
that
the amount of the irreversibility in the 
transition \(\rho \to \sigma \to \rho\), that is \( \rho \wobie \sigma \), will 
somehow acquire an extreme value when the target \(\sigma\) is a singlet.
But in this section, 
we will show by an explicit example that this is \emph{not} the case.  (See 
Refs. \cite{PlenioCirac,Acin} in this regard.)

First note that 
in the previous section, we have already shown that 
\[(\rho \wobie \sigma) < 
(\rho \wobie \psi^-) 
\]
holds for certain choices of the states \(\rho\) and \(\sigma\). More such examples will be exhibited 
in Section \ref{samemaxcorr}.

In this section, we will 
show that this inequality  
is \emph{not} true for all combinations of \(\rho\) and \(\sigma\).
We will present  cases where this inequality  is strictly reversed, that is where 
\begin{equation}
\label{reverse}
(\rho \wobie \sigma) > (\rho \wobie \psi^{-})
 \end{equation}
is true.

%Actually, we note that the inequality (\ref{main}) 
%is \emph{not} true for all combinations of \(\rho\) and \(\sigma\).
Indeed for \emph{arbitrary} states \(\rho\) and \(\sigma\), 
\((\rho \wobie \sigma) = 1\) for 
\(\rho = \sigma\), while 
\[(\rho \wobie \psi^{-}) = R(\rho \to \psi^{-})R(\psi^{-} \to \rho),\]
 being 
the rate at which we get back the state \(\rho\) in  a return journey via the target 
state \(\psi^{-}\), must  remain \(\leq 1\). This is just another way of stating 
that the entanglement cost of a state cannot be less than its distillable entanglement.
Consequently, 
we have 
\[(\rho \wobie \sigma) \geq (\rho \wobie \psi^{-})\]
%the equation (\ref{reverse}) is satisfied 
whenever \(\rho = \sigma\). 
%Note that
%for \(\rho\) and \(\sigma\) to be a  Bell mixtures, (\ref{reverse}) is also true, that is \(R_{Diff}\) is 
%also negative for \(p=q\) as we have already mentioned in the section \ref{Bellmixtures}. 
We will also have 
a strict inequality, that is the inequality (\ref{reverse}) will hold (for \(\rho = \sigma\)),
%\(R(\rho \rightarrow \sigma)R(\sigma \rightarrow \rho)\) strictly greater
%than  
%\(R(\rho \rightarrow \psi^{-})R(\psi^{-} \rightarrow \rho)\)
once we have \(\rho \wobie \psi^{-}\) 
strictly less than unity, for some \(\rho\). 
It may seem that this is true for any nondistillable state, i.e., 
states \(\rho\) for which \(R(\rho \to \psi^-) =0\) 
(this includes separable as well as bound entangled states \cite{PRLPLA}).
However for  the right hand side of the inequality (\ref{reverse}) to vanish for any nondistillable state \(\rho\),
one must also have finite \(R(\psi^- \to \rho)\). If  \(R(\psi^- \to \rho)\) is 
arbitrarily high (as we know is true at least for any separable state), one must consider 
some kind of limiting procedure. It is a nontrivial question as to what limiting procedure
one must consider in such a case. We will come back to this question in Section \ref{conclu}. 
%while the left hand side is of course unity, whenever \(\rho = \sigma\).
However for some bound entangled states \(\rho\), it was shown that  \(R(\psi^- \to \rho)\) 
is finite so that the inequality (\ref{reverse}) is satisfied for such states (the left hand side being unity
and the right hand side vanishing) \cite{irrev}. 
The inequality (\ref{reverse}) is satisfied even for some 
distillable states (i.e.  states \(\rho\) for which \(R(\rho \to \psi^-) >0\))
as was shown in Refs. \cite{irrev, DVC}.

\subsection{A case study: \(\rho\) and \(\sigma\) are mixtures of two Bell states}
Let us take  \(\rho\) and \(\sigma\) as in eqs. (\ref{rhotwoBell}) and (\ref{sigmatwoBell}).
It was shown in Ref. \cite{DVC} that 
\(D(\rho)\) is strictly less than \(F(\rho)\)
for \(1/2 < p < 1\). Consequently
we have 
\[(\rho \wobie \sigma) > 
(\rho \wobie \psi^-) 
\]
% is \emph{strictly} reversed 
whenever \(p = q \ne 1/2, 1\). 
The opposite inequality holds, i.e.
\[(\rho \wobie \sigma) < 
(\rho \wobie \psi^-) 
\] 
is true for any \(q\) and a sufficiently high \(p\), as was shown in the previous section (see also 
Fig \ref{nonzerof}).
Therefore it seems that with respect to the transition rates, maximally entangled states 
do not have a special status. Related points were made in Refs. \cite{PlenioCirac, Acin}. However
in their cases, the nonmaximally entangled ``extreme'' state was a nonmaximally entangled \emph{pure} state
and the considerations were in the non-asymptotic regime.

To get a more clear picture of what 
is going on, 
let us try to estimate the behavior of the difference 
\begin{equation}
\label{diff_specific}
R_{Diff} =  
(\rho \wobie \psi^-) - (\rho \wobie \sigma) 
\end{equation}
in the case when 
\(\rho\) and \(\sigma\) are given by eqs. (\ref{rhotwoBell}) and (\ref{sigmatwoBell}),
for \(p\in(1/2, 1)\) with a \emph{fixed} \( q \neq 1/2, 1\). 
From Fig. 1, it is clear that \(R_{Diff}\) is positive for 
\(p \in (1 - \varepsilon, 1)\) for some \(1 - \varepsilon > q\).

We have already shown that there are states for which
\[(\rho \wobie \sigma) > 
(\rho \wobie \psi^-) 
\]
for \(\rho = \sigma\).
%will now show that there do exist states
%\(\rho\) and \(\sigma\), with \(\rho \ne \sigma\), for 
%which the inequality (\ref{reverse}) is satisfied.  
From the continuity of 
\(R_{Diff}\) (Section \ref{cont}) 
and the fact that the set of distillable states is an open set \cite{Borda}, it follows
that this inequality holds also for \emph{unequal} \(\rho\) and \(\sigma\).
% The proof of continuity also provides us an arbitrary  
%class of states for which the inequality (\ref{reverse}) is satisfied. 
The above argument using continuity shows that for some states,
 it is better \emph{not} to return back via the singlet but via some other states.

Coming back to estimating the behavior of \(R_{Diff}\)
in the case when 
\(\rho\) and \(\sigma\) are given by eqs. (\ref{rhotwoBell}) and (\ref{sigmatwoBell}),
for \(p\in(1/2, 1)\) with a \emph{fixed} \( q \neq 1/2, 1\), it follows that \(R_{Diff}\) has a negative 
value around the point \(p = q\). Note that we are using the fact that mixtures of two Bell states,
if not mixed in equal proportions (when it is separable), are distillable states and as the set of 
distillable states is an open set \cite{Borda}, the class of all mixtures (excepting equal mixture case)
of two Bell states are in an open set of distillable states,
 so that the considerations in Section \ref{cont}
become applicable.

\begin{figure}[tbp]
  \epsfig{figure=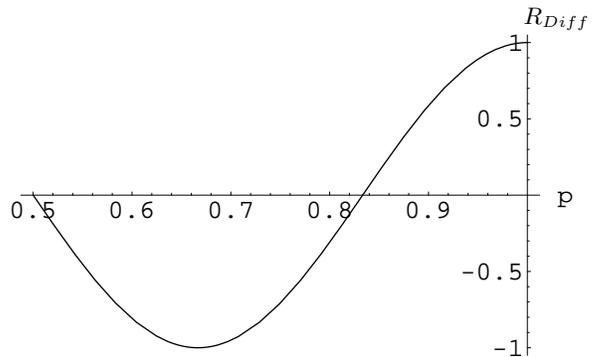,width=0.45\textwidth} 
\put(-28,127){\(R_{Diff}\)}
\caption{Hypothetical shape of
\(R_{Diff}\) in the case when 
\(\rho\) and \(\sigma\) are given by eqs. (\ref{rhotwoBell}) and (\ref{sigmatwoBell}),
for \(p\in(1/2, 1)\) with a \emph{fixed} \( q \neq 1/2, 1\).
We take \(q = 2/3\) in the figure.
Let us warn the reader that we compose facts and imagination in drawing the curve
(see text for full details). 
Notice that there is (at least) one point where \(R_{Diff}=0\). 
The set of pairs of states \(\{\rho, \sigma\}\), 
for which this happens,  probably has some special properties,
 as 
in the transformation \(\rho \wobie \sigma\) behaves like a transformation 
to a singlet. This  
is another way to see that, with respect to 
transformation rates, there is  nothing special about a singlet. 
} 
\label{rdifffig}
\end{figure}

In Fig \ref{rdifffig}, we try to plot \(R_{Diff}\) as a function of \(p\), in the case when 
\(\rho\) and \(\sigma\) are given by eqs. (\ref{rhotwoBell}) and (\ref{sigmatwoBell}),
for \(p\in(1/2, 1)\) with a \emph{fixed} \( q \neq 1/2, 1\).
In the figure, we take \(q = 2/3\).  All we know is that \(R_{Diff} = 1\) for 
\(p=1\) and \(R_{Diff} = -1\) for 
\(p=q\). Also we know that near the point \(p=1\), there is a certain neighborhood 
\((1-\varepsilon, 1)\), in which \(R_{Diff}\) is positive (see Fig. \ref{nonzerof}). 
Due to continuity, these two regions must meet. And in the process, \(R_{Diff}\) must cross the 
\(R_{Diff} = 0\) line at least once. We do not know if there are more than one crossing. 
It will be very interesting to find some general properties of the points 
where \(R_{Diff}\) vanishes. The set of all pairs of states \(\{\rho,\sigma\}\),
 for which 
\(R_{Diff}\) vanishes probably has some interesting properties, as 
the transformation \(\rho \wobie \sigma\) behaves like a transformation to a singlet. 
This is another way to see that maximally entangled states have no speciality 
with respect to transformation rates.
In the figure, we 
join the two portions near \(p = 1\) and near \(p = q\) by 
a monotonic curve. This monotonicity is by no means known. On the left of the 
point \(p = q (=2/3)\), we draw the \(R_{Diff}\) curve as monotonically reaching the value \(0\) as 
\(p \to 1/2\). Neither this monotonicity nor the limiting value are known. 
Note that for \(p = 1/2\), \(\rho\) is a separable state. We will come back to the issue
of the limiting value of the transformation rates, near a separable state, in the 
concluding section.

\section{Entanglement of formation for maximally correlated states}
\label{maxcorr}

The class of states 
for which we have proved that the inequality (eq. (\ref{main}))
\[(\rho \wobie \sigma) > (\rho \wobie \psi^{-})\]
holds (in  Section \ref{Bellmixtures}),  is a relatively small class of states. Hence 
the nonextremal nature of the singlet (in terms of asymptotic LOCC transformation rates) 
is depicted using this small class.
Can the same considerations be extended to a larger class? In the next Section, we try to make the extension
to the class of maximally correlated states \cite{Rainssemi}
\be
\rho^{mc} = \sum a_{ij}\left|ii\right\rangle\left\langle jj\right|.
\label{eq-maxcor}
\ee 
%We will do this extension in the next Section.

There are a 
number of obstacles in such an enterprise. We deal with these obstacles in this Section.
The distillable entanglement 
for the class of maximally correlated states is not known. The PPT-distillable entanglement (the 
optimal asymptotic fraction of faithful maximally entangled states obtainable by any 
superoperator preserving the positivity of partial transposition \cite{PeresHorodecki})
for any \(\rho^{mc}_{AB}\) from this class is known to be \cite{Rainssemi, huge} 
(see also \cite{WZ})
\[S(\rho^{mc}_{A}) - 
S(\rho^{mc}_{AB}),\] 
where \(S(\rho)\) is the von Neumann entropy of \(\rho\) and \(\rho^{mc}_{A} = 
\tr_{B}(\rho^{mc}_{AB})\).

It has been conjectured in \cite{Rainssemi} that the PPT-distillable entanglement 
(\(D_{\Gamma}\))  is 
equal to the distillable entanglement under LOCC operations for 
any \(\rho^{mc}\). 
In general it is not known whether there are states for 
which PPT-distillable entanglement and LOCC-distillable entanglement are 
provably different. In Section \ref{conclu}, we will  discuss this open problem from the 
perspective of the results obtained in this paper.   
Among the maximally correlated states, 
for mixtures of two Bell states as also for all pure states, these two quantities are equal.
This is also true for certain other states of the class of maximally correlated states, 
as has been checked in 
Refs. \cite{amader}. Our further steps may therefore be restricted by this problem.
Note however that to check whether the inequality (\ref{main}) is 
satisfied by using inequality (\ref{gerakol}), we may bypass this problem
by choosing \(\rho\) to be a state whose (LOCC-) distillable entanglement 
is known while \(\sigma\) to be a state whose PPT-distillable entanglement is known. 
This is due to the fact that the class of PPT superoperators is strictly larger
than the LOCC class (``strictly",  because of the existence of bound entangled states 
\cite{PRLPLA}). Consequently the (LOCC-) distillable entanglement \(D(\sigma)\)
of \(\sigma\) is smaller than or equal to its PPT-distillable entanglement 
\(D_{\Gamma}(\sigma)\). Thus we will have inequality (\ref{gerakol}) and 
hence inequality (\ref{main}), once we have 
\[\frac {D_{\Gamma}(\sigma)}{F(\sigma)} < \left(\frac {D(\rho)}{F(\rho)}\right)^2.\]

The next obstacle is that  entanglement cost is also not known 
for the maximally correlated states.  In general, entanglement cost is 
equal to  regularization of entanglement of formation $E_f$ \cite{HHT}. 
If, for  a given  state $\rho$ we have 
\[E_f(\rho\otimes\rho)=2E_f(\rho),\] 
then  the two quantities are equal.  Below we will show, on the lines of \cite{DVC} (see also \cite{Shor}), 
that the entanglement 
of formation is additive for maximally correlated states. Consequently,
the entanglement cost is obtained for these states in 
\(2 \otimes 2\), as entanglement of formation is known for all states of two qubits
\cite{Wootters}. 
 Also we supply an optimization procedure for calculating the entanglement 
of formation for maximally correlated states in 
\emph{arbitrary} dimensions. The procedure is simpler than that is  
contained in the very definition of entanglement of formation.
Having obtained the entanglement cost for maximally correlated states in \(2 \otimes 2\)
and as the PPT-distillable entanglement is known for all maximally correlated states \cite{Rainssemi, huge},
we show that all nonpure entangled maximally correlated states in \(2 \otimes 2\) have 
irreversibility in asymptotic LOCC manipulations of entanglement.

\subsection{Additivity of entanglement of formation for maximally correlated states in arbitrary dimensions}

In this subsection, we provide some formula for \(E_f\) 
for
maximally correlated states in \(d \otimes d\)
and 
we will show that their  entanglement of formation is additive.
Entanglement of formation \(E_f\) of a bipartite state \(\rho\) is defined as
\[
E_f(\rho)=\inf\sum_ip_i S_A(\psi_i)
\]
where $S_A(\psi)$ denotes entropy of reduction of the bipartite state $\psi$ to a single party and the 
optimization is taken over all decompositons 
$\rho=\sum_ip_i |\psi_i\>\<\psi_i|$
of $\rho$ 
into pure bipartite states. 
The maximally correlated states (\ref{eq-maxcor}) are determined 
by the matrix $a_{ij}$. One finds that the matrix has to be positive semidefinite
and of unit trace. Thus with any maximally correlated state $\rho^{mc}$ 
in \(C^d \otimes C^d\),
one can associate the state $\rho'$ of a \emph{single} system of the form
\be
\rho'=\sum_{ij}a_{ij} |i\>\<j|.
\ee
One notes that the support of $\rho^{mc}$ is spanned by the vectors of the form
\be
\psi=\sum_i c_i |i\>|i\>
\ee
where $\{ |i>\}$ is the same basis as the one used in the definition
of the maximally correlated state in eq.~(19).  As this is a subspace,
any decomposition of $\rho^{mc}$ will consist of vectors of the above
form.
%
%where $\{|i\>\}$ has to be the same basis as the one used in the 
%definition of the maximally correlated state in eq. (\ref{eq-maxcor}). Thus 
%any decomposition of $\rho^{mc}$ will consist of the vectors of the above form.
Consider then any decomposition $\{p_k,\psi_k\}$ with
$\psi_k=\sum_i c_i^k|i\>|i\>$. The coefficients have to satisfy 
the constraints
\[
\sum_kp_kc_i^k \overline{c_j^k}=a_{ij}.
\]
Treating $\{c_i^k\}_i$ as  vectors $x_k$ belonging to the
Hilbert space $C^d$ of a single system, we obtain 
\be
\sum_kp_k |x_k\>\<x_k| = \varrho'.
\ee
The entropy $S_A(\psi_k)$ is 
equal to Shannon entropy of the diagonal elements 
of the state $|x_k\>\<x_k|$ in the basis $|i\>$:
\be
S_A(\psi_k)=-\sum_i|\<i|x_k\>|^2 \log_{2} |\<i|x_k\>|^2\equiv H(x_k).
\ee
We then obtain the following formula for entanglement of formation
of the maximally correlated state $\rho^{mc}$: 
\be
E_f(\varrho_{mc})=\inf \sum_k p_k H(x_k)
\ee
where the infimum is taken over all decompositions of 
the state $\rho'$ (defined above) into pure states $x_k$.
Similarly as in the definition of entanglement of formation, 
we can take infimum over all decompositions of $\rho'$ 
(including mixed states as members of decomposition). 
The obtained formula is simpler than the original optimization 
procedure because it  involves state of  one system 
(not a compound one).

Let us now pass to the problem of whether $E_f=F$ for 
maximally correlated states.  In Ref. \cite{DVC} it was shown that 
any state with support within subspace $V\subset \hcal_A\otimes \hcal_B$ 
has $E_f=F$ if the subspace has the following property: the map 
$\Lambda: \bcal(V)\to \bcal(\hcal_A\otimes \hcal_B)$
given by partial trace is so-called entanglement breaking map 
(using such a map as a channel, one cannot share an entangled state). 
By extending Example 1 of Ref. \cite{DVC} one easily finds that the 
subspace spanned by vectors of the form $|i\>|i\>$ has such a property.
The class of states with support lying within the supspace 
coincides with maximally correlated ones, so that $F=E_f$ 
for those states. 

\subsection{Irreversibility in asymptotic manipulations of entanglement 
for maximally correlated states in \(2\otimes 2\)}

\begin{figure}[tbp]
  \epsfig{figure=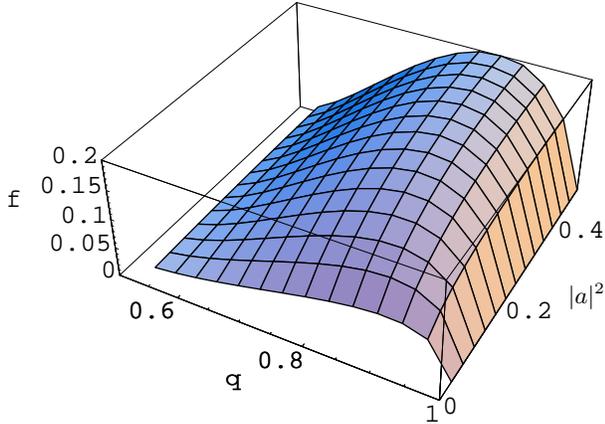,width=0.45\textwidth} 
\put(-15,60){\(|a|^2\)}
\caption{Plot for \( f = F - D_\Gamma\) on the \((q,|a|^2)\)-plane
 for maximally correlated states \(\sigma\) given by 
(\ref{maxcor2x2}) in \( 2 \otimes 2\).} 
\label{maxcorrfig}
\end{figure}

In the previous subsection, we have shown that entanglement of formation is additive for 
maximally correlated states  (given by eq. (\ref{eq-maxcor})) in \(d \otimes d\). 
%But we were not able to find the explicit formula for
%entanglement of formation or cost. 
As entanglement of formation is known for all two qubit states \cite{Wootters},
we therefore are able to calculate the entanglement cost \(F\) for 
all maximally correlated states of two qubits. Any such state in \(2 \otimes 2\) can be written 
as 
\begin{equation}
\label{maxcor2x2}
\sigma =  (1-q)\left|\phi\right\rangle\left\langle \phi \right|
+ q \left|\psi \right\rangle\left\langle \psi \right|, \quad q\in[\frac{1}{2},1]
\end{equation}
where
\[\left|\phi\right\rangle = a\left|00\right\rangle + b\left|11\right\rangle, \quad
\left|\psi\right\rangle = \bar{b}\left|00\right\rangle - \bar{a}\left|11\right\rangle,\]
with \(|a|^2 + |b|^2 = 1\).
Consequently, one finds that \cite{Wootters}
\[ F = h\left(\frac{1}{2} +  \frac{1}{2}\sqrt {1 - 4 (2q-1)^2 |a|^2|b|^2}\right)\]
for the maximally correlated state \(\sigma\) (given by eq. (\ref{maxcor2x2})), where
\(h(x) = -x \log_2 x - ( 1- x) \log_2 (1-x)\) is the binary entropy function.
 Now the PPT-distillable entanglement
\(D_\Gamma\) of such states is also known, as we noted earlier. In Fig. \ref{maxcorrfig}, we plot
\(F - D_\Gamma\) for \(\sigma\) on the \((q, |a|^2)\)-plane. 
We see that the value of \(F - D_\Gamma\) is strictly positive for all 
\(q \in (1/2, 1)\) and \(|a|^2 \in (0, 1/2]\). As \(D_\Gamma\) is greater than
or equal to \(D\), 
\(F - D\) is also strictly positive for these ranges. We therefore have irreversibility
in asymptotic manipulations of entanglement (i.e. \(D <F\))  for 
all nonpure entangled maximally correlated states in 
\(2 \otimes 2\). 

\section{More cases to show that the singlet is not special with respect to asymptotic LOCC 
transformation rates
%More cases in which eq. (\ref{main}) is satisfied
%Irreversibility in \(\rho \wobie \sigma\) when \(\rho\) is maximally correlated state
}
\label{samemaxcorr}

Having calculated the entanglement cost of the maximally correlated
states in \(2 \otimes 2\), we will now be able to find more examples where the inequality (eq. (\ref{main}))
\[\rho \wobie \sigma < 
\rho \wobie \psi^{-}\] 
is satisfied.   
Let us consider the case when \emph{\(\rho\) is a mixture of two Bell states  and \(\sigma\) is a maximally 
correlated state of two qubits.} So let \(\rho\) be given by eq. (\ref{rhotwoBell}), i.e.
\[
\rho = (1-p)\left|\phi^{+}\right\rangle\left\langle \phi^{+} \right|
+ p \left|\phi^{-}\right\rangle\left\langle \phi^{-} \right|\]
and let 
\(\sigma\) be given by (\ref{maxcor2x2}), i.e.
\[\sigma =  (1-q)\left|\phi\right\rangle\left\langle \phi \right|
+ q \left|\psi \right\rangle\left\langle \psi \right|, \quad q\in[\frac{1}{2},1]
\]
where
\[\left|\phi\right\rangle = a\left|00\right\rangle + b\left|11\right\rangle, \quad
\left|\psi\right\rangle = \bar{b}\left|00\right\rangle - \bar{a}\left|11\right\rangle,\]
with \(|a|^2 + |b|^2 = 1\).

As is easily checked, for every value of \(\left|a\right|\), we get a 
similar surface 
for the function 
\(D^2(\rho)F(\sigma) - F^2(\rho)D(\sigma)\) 
over the \((p, q)\) -plane as in Fig. \ref{nonzerof}. Therefore for a fixed \(q\) (and 
\(\left|a\right|\)), 
there is always a region \((1-\varepsilon, 1)\) (with \(1 - \varepsilon <q\)) for which 
\(\rho\) (for \(p\in(1-\varepsilon, 1)\)) and \(\sigma\) satisfies the inequality (\ref{main}).
That is, \(R_{Diff}\) is positive for \(p\in(1-\varepsilon, 1)\) (for some 
\(1-\varepsilon < q \)),
for all fixed values 
of \(q\) and \(\left|a\right|\).

We have \(\rho = \sigma\) on the intersection of \(\left|a\right| = 1/\sqrt{2}\)
and \(p=q\), whereby \(R_{Diff}\) is negative on that intersection. Via continuity of
\(R_{Diff}\) as well as 
due to the fact that the set of distillable states is open \cite{Borda},
\(R_{Diff}\) will be negative for unequal \(\rho\) and \(\sigma\) also,
provided they are sufficiently close to each other as well as to the intersection of 
\(\left|a\right| = 1/\sqrt{2}\) and \(p=q\).

Exactly similar results are obtained even when both \(\rho\) and \(\sigma\)
are from the class of maximally correlated states, if we accept the conjecture
of Ref. \cite{Rainssemi} discussed above.

\section{Ratio Problem}
\label{ratio}

In this section, we will  view our results from 
the perspective of the ratio problem of entanglement 
measures \cite{Nielsen}.
Let us first briefly recall what is already known about the problem.
This will provide 
a better setting for the aspect of the problem that we want to discuss.
The distillable entanglement \(D(\rho)\) of a bipartite state \(\rho\) is defined 
as an (optimal) asymptotic \emph{ratio}. It is the optimal asymptotic fraction 
of the number of faithful singlets (\(\psi^{-}\))  obtainable 
via an LOCC protocol. It is therefore 
defined with the maximally entangled states as unit. As distillable entanglement is 
a measure of a physical quantity (indicating the potential 
of \(\rho\) to teleport, for example), the ratio of the distillable entanglements
of different states,  may be hoped to be independent of the chosen unit. The heights
of two persons must have the same ratio whether their heights are measured in 
centimetres or inches. However it turned out that 
it is not true. In Ref. \cite{Nielsen}, examples were cited for which
\[
\frac {D_{\sigma}(\rho_{1})}{D_{\sigma}(\rho_{2})} \ne 
\frac {D(\rho_{1})}{D(\rho_{2})}.
\]
The same problem arises for entanglement cost as well.

Let us now consider a complementary aspect of the ratio problem. 
The ratios discussed in Ref. \cite{Nielsen} were between different states.
Can we not have such a ratio for a single state? It should not be 
``different" properties of the same state. The height to weight ratio
of a person may vary in different unit systems. 
But the ratio of the height of an individual to the length 
of her/his arm must remain the same in any unit system. Is this 
the case with distillable entanglement and entanglement 
cost of a state? That is, are the ratios 
\(
\frac {D_{\sigma}(\rho)}{F_{\sigma}(\rho)}\) and  
\(
\frac {D(\rho)}{F(\rho)}
\) equal?

By what we have already shown in the previous sections, these ratios can be shown
to be unequal in certain cases. Indeed in the above case-studies
one needs to substitute \(R(\rho \rightarrow \sigma)\) by
\(D_{\sigma}(\rho)\) and \(R(\sigma \rightarrow \rho)\) by
\(\frac {1}{F_{\sigma}(\rho)}\) as also
\(R(\rho \rightarrow \psi^{-})\) by
\(D(\rho)\) and \(R(\psi^{-} \rightarrow \rho)\) by \(\frac {1}{F(\rho)}\)  
to obtain the envisioned nonequality 
\[\frac {D_{\sigma}(\rho)}{F_{\sigma}(\rho)}
\ne
\frac {D(\rho)}{F(\rho)}. 
\]

\section{Discussions}
\label{conclu}

We have shown that  the optimal asymptotic rate 
at which \(\rho\) is retrieved (under LOCC) in a cycle \(\rho \rightarrow \rho\)
via \(\sigma\) can be different for different \(\sigma\).  
This gives a new dimension to the 
fundamental irreversibility in asymptotic manipulations of entanglement
in that it provides in a natural way, a notion of the
``amount'' of the irreversibility in local asymptotic manipulations 
of a state.

For the considerations made in this paper, we take 
\(\rho\) and \(\sigma\) as two mixtures of two Bell states 
as also \(\rho\) as a mixture of two Bell states and \(\sigma\) as a 
maximally correlated state. 
%We have also shown the similar characteristics of 
%\(\rho \wobie \sigma \) taking  \(\rho\) as maximally correlated states. 

The considerations naturally led us to consider the quantity
\[R_{Diff} =  
(\rho \wobie \psi^{-}) -
(\rho \wobie \sigma) 
\]
whose behavior we have tried to judge in certain simple cases.
To judge the character of \(R_{Diff}\),
we have used either certain bounds on the transformation (derived in Section \ref{def}) or 
we have used certain continuity arguments (Section \ref{cont}). The exact value of the
quantity is not known for a single nontrivial case. 
Nevertheless,  one can 
have a feel for the general behavior of this difference.
Consider a distillable state \(\eta\) in \(m\otimes n\) for which 
\(D(\eta)\) is strictly less than \(F(\eta)\). 
%Let \(\psi^{-}\) be the 
%corresponding maximally entangled state. 
Then
\[
(\eta \wobie \psi^{-}) - 1 \equiv 
\frac{D(\eta)}{F(\eta)} - 1 < 0
\] 
Therefore \(R_{Diff}\),
for \(\rho = \sigma = \eta\), is strictly less than zero. 
From the continuity of \(R_{Diff}\), which we proved earlier,
and as the set of distillable states is open \cite{Borda},
 it follows that negativity (of \(R_{Diff}\)) remains for \(\rho\) and \(\sigma\)
sufficiently close to \(\eta\). 
On the other hand, 
\[
1 - (\psi^{-} \wobie \eta) \equiv 
1 - \frac{D(\eta)}{F(\eta)} > 0
\] 
Therefore \(R_{Diff}\)
is strictly positive for \(\rho = \psi^{-}\) and \(\sigma = \eta\). And again by continuity,
\(R_{Diff}\) is positive for \(\rho\) and \(\sigma\) sufficiently close to 
\(\psi^{-}\) and \(\eta\) respectively.

We tried to plot \(R_{Diff}\) in a simple case in Fig. \ref{rdifffig}. Related to this, there are 
some interesting open questions which we have discussed in Section \ref{maxnotspecial}.

Here we will like to discuss the issue of 
the behaviour of the quantity \(R_{Diff}\) near separable states.
For example, consider the case when \(\rho\) and \(\sigma\) are mixtures of two Bell states (given by 
eqs. (\ref{rhotwoBell}) and (\ref{sigmatwoBell})). Take a \emph{fixed} \(q \ne 1/2, 0\). 
We are interested to find the value of 
\(R_{Diff}\) as we approach the point   \(p= 1/2\). Note that at  \(p = 1/2\), \(\rho\) is a separable state.
It may seem that
in general, the rate \(\rho \wobie \sigma\) at which \(\rho\) is retrieved in a return journey via \(\sigma\)
is vanishing in the limit when  \(\sigma\) approaches to a separable state while \(\rho\) remains distillable.
In the case when \(\rho\) is given by
eqs. (\ref{rhotwoBell}), we have checked that \(\rho \wobie \psi^- \to 0\) as \(p \to 1/2\) \cite{DVC}.
In Fig. \ref{rdifffig}, we have  plotted \(R_{Diff} \to 0\) as \(p \to 1/2\)
with this intuition.

We will now see that if we assume that \(\rho \wobie \sigma \to 0\)  as \(\sigma\) approaches to a separable 
state (with \(\rho\) remaining distillable), then one can arrive at examples of states for which the 
PPT-distillable entanglement is strictly  greater than its LOCC distillable entanglement.
%One may ask here  
%another question about the value of the transformation 
%\(\rho \wobie \sigma \) in the case of separable states.
%We believe that  the value of that transformation will be 
%vanishing in that case as from separable states, 
%it is impossible to create the maximally enatangled state. 
%But this belief will give some opposite 
The example is against  the conjecture given in \cite{Rainssemi} that for maximally 
correlated states, \(D_{\Gamma} = D\). Let us 
mention however that we are not in a position to give a counterexample to this conjecture.
We merely show that a counterexample exists if we believe that \(\rho \wobie \sigma \to 0\) as 
\(\sigma\) approaches to a separable state (with \(\rho\) remaining distillable). 
Consider  the state 
\[ \rho(p) = p \left|00\right\rangle \left\langle 00 \right| + 
(1- p) \left|\phi^{+}\right\rangle \left\langle \phi^{+} \right|\]
and  the rate
\[\rho \wobie \psi^{-} = \frac{D(\rho)}{F(\rho)}.\] 
%\leq  \frac{D^{PPT}(\rho)}{F(\sigma)}. \]
Note here that the state is product when \(p = 1\). Therefore according to our assumption,
\(\frac{D(\rho)}{F(\rho)} \to 0\) as \( p \to 1\). 
For the state \(\rho\), we know the values of PPT-distillable entanglement \cite{Rainssemi, huge, WZ}
as well as its entanglement cost (Section \ref{maxcorr}). One can easily check that the quantity 
\(\frac{D_{\Gamma}(\rho)}{F(\rho)}\) tends to \(1/2\) as \(p\) tends to \(1\). 
Thus, modulo our assumption, we have that \(D_{\Gamma}\) is strictly greater than \(D\)
for states \(\rho(p)\) which are sufficiently close to \(\rho(1)\).

\begin{acknowledgments}
AS and US thanks Michael A. Nielsen for commenting on an earlier manuscript.
They also thank  Martin B. Plenio and Vlatko Vedral for discussions  during
their ESF Short Scientific visit to QOLS, Imperial College, London. 
This work is supported by the University of Gda\'{n}sk, 
Grant No. BW/5400-5-0236-2 and by the European Community under 
project EQUIP, Contract No. IST-11053-1999. 
\end{acknowledgments}


\begin{thebibliography}{9}
\bibitem{book1}{\small G. Alber, T. Beth, M. Horodecki, P. Horodecki, R. Horodecki,
M. R\"{o}tteler, H. Weinfurter, R. Werner and A. Zeilinger,} \emph{\small Quantum
Information: An Introduction to Basic Theoretical Concepts and Experiments}{\small ,
Springer Tracts in Modern Physics 173} {\small (Springer-Verlag,
Berlin, 2001).
\bibitem{book2}M.A. Nielsen and I.L. Chuang, \emph{Quantum Computation and Quantum 
Information}, Cambridge University Press, Cambridge, 2000.}{\small \par}
\bibitem{tele}{\small C. H. Bennett, G. Brassard, C. Cr\'epeau, R. Jozsa, A. Peres,
and W. K. Wootters, Phys. Rev. Lett.} \textbf{\small 70},
{\small 1895 (1993).}{\small \par}
\bibitem{tensorname}We denote \(\mathbb{C}^m \otimes \mathbb{C}^n\) by \(m \otimes n\), where \(\mathbb{C}^r\)
is the \(r\)-dimensional complex Hilbert space.
\bibitem{2x2}We remember that maximally entangled states in any dimensions (as also any pure state)
are reversibly transformable by LOCC to a maximally entangled state in \(2 \otimes 2\) \cite{concpure}.
\bibitem{concpure}C.H. Bennett, H.J. Bernstein, S. Popescu, and B. Schumacher, 
Phys. Rev. A, \textbf{53}, 2046 (1996) (quant-ph/9511030).
\bibitem{huge}{\small C. H. Bennett, D. P. DiVincenzo, J. A. Smolin, and 
W. K. Wootters, Phys. Rev. A {\bf 54}, 3824 (1996) (quant-ph/9604024).}{\small\par}
\bibitem{Rains}{\small E.M. Rains, Phys. Rev. A \textbf{60}, 173 (1999) 
(quant-ph/9809078).}{\small\par}
\bibitem{HHT}{\small P.M. Hayden, M. Horodecki, and B.M. Terhal, J. Phys. A: Math. Gen. 
{\bf34(35)}, 6891 (2001) (quant-ph/0008134). }{\small\par}
\bibitem{irrev}{\small G. Vidal and J. I. Cirac,} \emph{\small Phys. Rev. Lett.}
\textbf{\small 86}{\small , 5803 (2001) (quant-ph/0102036); G. Vidal
and J. I. Cirac,} \emph{\small When only two thirds of entanglement
can be distilled}{\small , quant-ph/0107051.}{\small \par}
\bibitem{DVC}{\small G. Vidal, W. Dur, and J.I. Cirac, \emph{Entanglement cost
 of mixed states}, quant-ph/0112131.}{\small\par}
\bibitem{bound}{\small The intuitive feeling for the existence
of such states arose when it was shown that there exist entangled states for which the 
distillable entanglement is vanishing \cite{PRLPLA}.}{\small\par}
\bibitem{PRLPLA}{\small M. Horodecki, P. Horodecki, and R. Horodecki, Phys. Rev. Lett.}
\textbf{\small 80} {\small 5239 (1998) (quant-ph/9801069);} 
{\small  P. Horodecki, Phys. Lett. A}
\textbf{\small 232} {\small 333 (1997) (quant-ph/9703004).}{\small\par} 
\bibitem{Vidalcont}G. Vidal, \emph{On the continuity of 
asymptotic measures of entanglement}, quant-ph/0203107.
\bibitem{Shor}{\small P.W. Shor, \emph{Additivity of the Classical Capacity of 
Entanglement-Breaking Quantum
Channels}, quant-ph/0201149.}{\small\par}
\bibitem{Wootters}{\small W.K. Wootters, Phys. Rev. Lett. \textbf{80}, 2245 (1998) 
(quant-ph/9709029).}{\small\par}
\bibitem{Nielsen}{\small M.A. Nielsen, \emph{On the units of bipartite entanglement: 
Is sixteen ounces of entanglement always equal to one pound?}, quant-ph/0011063.}{\small\par}
\bibitem{PeresHorodecki}{\small A. Peres, Phys. Rev. Lett.
\textbf{77}, 1413 (1996) (quant-ph/9604005);
M. Horodecki, P. Horodecki, and R. Horodecki, Phys. Lett. A \textbf{223}, 1 (1996) 
(quant-ph/9605038).}{\small\par}
\bibitem{Rainssemi}{\small E.M. Rains, 
Phys. Rev. A \textbf{60}, 179 (1999) (quant-ph/9809082);
Phys. Rev. A \textbf{63}, 019902 (2001); \emph{A semidefinite 
program for distillable entanglement}, quant-ph/0008047.}{\small\par}
\bibitem{miary}
{M. Horodecki, P. Horodecki, and R. Horodecki,}
{Phys. Rev. Lett.} \textbf{84},
2014 %-2017
(2000)
(quant-ph/9908065).
\bibitem{Donald}
M. Donald, M. Horodecki, and O. Rudolph, J. Math. Phys. \textbf{43}, 4252
%-4272 
(2002) (quant-ph/0105017).
\bibitem{review} M. Horodecki, QIC {\bf 1}, 3 (2001).
\bibitem{Plenio}
V. Vedral and M.B. Plenio,
{Phys. Rev. A} \textbf{57}, 1619 
%-1633
(1998)
(quant-ph/9707035).
\bibitem{PlenioCirac} S.F Huelga, C. Macchiavello, T. Pellizzari, A.K. Ekert, M. B. Plenio, and J.I. Cirac,
Phys.Rev.Lett. \textbf{79} 3865 (1997) (quant-ph/9707014).
\bibitem{Acin}A. Ac\'\i n, T. Durt, N. Gisin, and J.I. Latorre, Phys. Rev. A \textbf{65}, 052325 (2002)
(quant-ph/0111143).
\bibitem{Borda}P. Horodecki, J.I. Cirac, and M. Lewenstein, 
\emph{Bound entanglement for continuous variables is a rare phenomenon}, quant-ph/0103076.
\bibitem{WZ}S. Yu and Y. Zhang, \emph{Calculating the relative entropy of entanglement}, quant-ph/0004018.
\bibitem{amader}{\small S. Ghosh, G. Kar, A. Roy, A. Sen(De), and U. Sen, Phys. Rev. Lett
{\textbf 87}, 277902 (2001) (quant-ph/0106143); Y.-X. Chen and D. Yang, 
\emph{The relative entanglement of Schmidt correlated states and distillation},
quant-ph/0204152.}{\small\par}



\end{thebibliography}
\end{document}